\documentclass[lineno]{svproc}
\usepackage[utf8]{inputenc}
\usepackage{graphicx}%
\usepackage{multirow}%
\usepackage{amsmath,amssymb,amsfonts}%
\usepackage{mathrsfs}%
\usepackage[title]{appendix}%
\usepackage{xcolor}%
\usepackage{textcomp}%
\usepackage{manyfoot}%
\usepackage{booktabs}%
\usepackage{algorithm}%
\usepackage{algorithmicx}%
\usepackage{algpseudocode}%
\usepackage{listings}%
\usepackage{lineno} 
\usepackage{amssymb}
\usepackage{url}
\usepackage[skip=4pt]{caption}
\usepackage{xcolor} 

\begin{document}
\mainmatter              
%
\title{Enhanced Cyber Threat Intelligence by Network Forensic Analysis for Ransomware as a Service(RaaS) Malwares }
\titlerunning{RaaS}  
%
\author{Sharmila S P\inst{1,}\inst{2}, 
Aruna Tiwari\inst{1}, Narendra S Chaudhari\inst{3} }
%
%
\authorrunning{Sharmila et al.} 
%
%
\institute{Indian Institute of Technology Indore, Madhya Pradesh, India\\
\and
Siddaganga Institute of Technology, Tumakuru, Karnataka, India\\ 
\and 
Assam Science and Technology University, Jalukbari, Guwahati, Assam, India
\email{\{phd2201101012,artiwari\}@iiti.ac.in,sharmila@sit.ac.in,vc@astu.ac.in}
}

\maketitle              
\begin{abstract}
In the current era of interconnected cyberspace, there is an adverse effect of ransomware on individuals, startups, and large companies. Cybercriminals hold digital assets till the demand for payment is made. The success of ransomware upsurged with the introduction of Ransomware as a Service(RaaS) franchise in the darknet market. Obfuscation and polymorphic nature of malware make them more difficult to identify by Antivirus system. Signature based intrusion detection is still on role suffering from the scarcity of RaaS packet signatures. We have analysed RaaS samples by network forensic approach 
to investigate on packet captures of benign and malicious network traffic. The behavior analysis of RaaS family Ransomwares, Ryuk and Gandcrab have been investigated to classify the packets as suspicious, malicious, and non-malicious which further aid in generating RaaS packet signatures for early detection and mitigation of ransomwares belonging to RaaS family. More than 40\% of packets are found malicious in this experiment. The proposed method is also verified by Virus Total API Approach. Further, the proposed approach is recommended for integration into honeypots in the present scenario to combat with data scarcity concerned with malware samples(RaaS). This data will be helpful in developing AI-based threat intelligence mechanisms. In turn enhance detection, prevention of threats,  incident response and risk assessment. 
\end{abstract}
\section{Introduction}
\small
Ransomware as a Service (RaaS) represents a disturbing evolution in the realm of cybercrime, where malicious actors offer ready-to-use ransomware packages to affiliates, enabling them to conduct extortion campaigns with relative ease and minimal technical expertise\cite{meland2020ransomware}. This business model has fueled a surge in ransomware attacks globally, targeting organizations of all sizes across various industries. Distributed ledgers, blockchain \cite{sharmila2021operative} and interplanetary file systems are \cite{karapapas2020ransomware} offering more robust services with this model for launching cybercrimes.
\noindent The RaaS ecosystem \cite{meland2020ransomware} \cite{greenstein2022impact} \cite{ruellan2023conti},   comprises  the following players:

\noindent \textbf{1. Developers and Operators:} The individuals or groups responsible for creating and maintaining ransomware variants handling malware development, infrastructure management, and customer support. Often advertising their services on dark web forums, attracting potential affiliates.

\noindent \textbf{2. Affiliates:} The customers who purchase ransomware packages from developers for  distributing them to target victims. They are experienced cybercriminals looking to expand their operations or individuals seeking quick profits through cyber extortion, receiving a percentage of the ransom payments collected from victims.

\noindent \textbf{3. Payment Processors:} Facilitators for collection and laundering of ransom payments on behalf of ransomware operators, providing anonymous payment channels\cite{chris2024guide}, such as cryptocurrency exchanges or money mule networks, to receive and transfer ransom funds securely.

\noindent \textbf{4. Support Services:} Group of RaaS operators offer support services to affiliates, including technical assistance, troubleshooting, and advice on maximizing ransom payments, with streamlining operations to ensure the success of ransomware campaigns.\\
This ecosystem democratizes cybercrime by lowering the barriers to open entry for attackers. Affiliates can access sophisticated ransomware tools and infrastructure without needing to develop malware from scratch or maintain complex command-and-control (C2) systems\cite{hull2019ransomware}. As a result, RaaS has become a lucrative venture for both developers and affiliates, leading to a proliferation of ransomware attacks worldwide. Some notable RaaS platforms till 2025 are Darkside\footnote{https://www.state.gov/darkside-ransomware-as-a-service-raas}, REvil\footnote{https://www.state.gov/transnational-organized-crime-rewards-program-2/sodinokibi-ransomware-as-a-service-raas/}, Conti\footnote{https://www.cisa.gov/news-events/alerts/2021/09/22/conti-ransomware}, Maze\footnote{https://www.cloudflare.com/learning/security/ransomware/maze-ransomware/}, Lockbit\footnote{https://www.mcafee.com/blogs/other-blogs/mcafee-labs/tales-from-the-trenches-a-lockbit-ransomware-story/}, NetWalker\footnote{https://www.mcafee.com/blogs/other-blogs/mcafee-labs/take-a-netwalk-on-the-wild-side/}, Avaddon\footnote{https://malpedia.caad.fkie.fraunhofer.de/details/win.avaddon}, Eggregor\footnote{https://heimdalsecurity.com/blog/egregor-ransomware/}. Some RaaS Groups supplying ransomwares as per 2025 statistics are:
Hive, Darkside, Revil, Dharma, Lockbit, Locky, Goliath, Stampado, Encryptor, Ragnarok, Prolock, Crylock, Nefilin. Apart from these, the major gap identified is the scarcity of RaaS packet signatures for signature-based IDS which are still on role in network appliances. \\
Following are the significant challenges posed by RaaS for a profound impact on cybersecurity landscape, law enforcement\cite{darkweb2023trendmicro}, and security professionals\cite{payne2021multiple} \cite{karapapas2020ransomware}.\\
\noindent \textbf{C1. }Frequency of Attacks: The rise of RaaS platforms has fueled a global increase in ransomware attacks across various sectors, including healthcare, education, finance, and government.

\noindent \textbf{C2. }Sophistication of Attacks: RaaS operators constantly enhance their malware to avoid detection and enhance infection rates, leading to stealthier and more advanced ransomware strains that can bypass conventional security defenses.

\noindent \textbf{C3. }Financial losses and Operational disruption: Ransomware attacks inflict financial losses, disrupt operations, and lead to reputational damage and legal liabilities for organizations.

\noindent \textbf{C4. }Challenges for Law Enforcement: The decentralized nature of RaaS operations complicates law enforcement efforts. Anonymous RaaS operators on the dark web pose attribution challenges.

RaaS represents a paradigm shift in cybercrime, empowering individuals and groups to engage in lucrative extortion\cite{meurs2024deception} schemes with minimal effort. The widespread availability of ransomware tools and infrastructure poses significant challenges for organizations seeking to defend against these threats. Effective mitigation strategies require a multi-faceted approach, including robust cybersecurity defenses, AI based Intrusion Prevention mechanisms\cite{10.1007/978-981-19-2500-9_47}, \cite{sharmila2022distinguished}, threat intelligence sharing, and international cooperation\cite{iu2024trans} to combat this growing menace.\\
The widespread availability of ransomware tools and infrastructure rises significant challenges for organizations seeking to defend against these embedded threats\cite{sharmila2025unveiling}. 
Effective mitigation strategies require a multi-faceted approach, including robust cybersecurity defenses, AI based Intrusion Prevention mechanisms\cite{10.1007/978-981-19-2500-9_47,sharmila2022distinguished}, threat intelligence sharing, and international cooperation to combat this growing menace.
In this paper we address the problem by presenting a hybrid approach of network forensics integrated with VirusTotal to extract RaaS behavior to frame packet signature, suggesting a reverse engineering tactic to find hidden files concerned with its network traffic. Further investigate on these files to explore multiple attack vectors. Identifying malicious packets can reveal insights into RaaS attacker tactics, techniques, and procedures (TTPs), which can be used to improve threat intelligence and incident response. Ryuk and Gandcrab samples are considered for our experiments at present. Proposed classification results can be used to automate security controls, to block or alert on similar malicious traffic in the future. Analyzing RaaS traffic can help incident responders understand the scope and impact of a security breach. Identifying suspicious or malicious packets can aid in network traffic analysis, helping security teams to identify potential security gaps or vulnerabilities.\\

\noindent\textbf{The main contributions of this work are:}
\begin{enumerate}
	\item Ransomware samples of RaaS family are collected and executed in an isolated vulnerable honeypot with continuous monitoring to study the behavior of ransomware through network traffic, API calls, memory dumps etc.
	\item Packets captured by network traffic analysis during ransomware execution, are further analysed to investigate on its malicious activities and behavior.
	\item Classify the packets into suspicious, malicious and non-malicious for Feature set extraction and threat identification, verifying the same with VirusTotal. 
	
\end{enumerate}
The rest of the paper is organized as follows. Initially in section \ref{Background}, Background and related work are discussed, next in section \ref{Proposed}, detailed description of the proposed work is presented, further results are portrayed in section \ref{Results}. Finally it is concluded in section \ref{Conclusion}, with a glimpse of future work. 
\section{Related Work}
\label{Background}
\small
Studying the behavior of ransomware is very challenging because Command and Control(C2) servers are taken down when ransomware sample is deactivated, this makes it impossible to track the behavior\cite{berrueta2020open} by new detection tools. Apart from identifying malware family\cite{sharmila2025leveraging}, studying the behavior of ransomware is also challenging because C2 servers are taken down when ransomware sample is deactivated, this makes it impossible to track the behavior by new detection tools.
Collection of samples is equally challenging with development of new algorithms for detection of malwares. Two promising solution for the challenges connected with ransomware detection are i)either observe the network traffic or ii)observe the ransomware activity on files. Apart from traffic capturing, data handling is easier when IO operations on user documents by the infected host is studied. It is necessary to extract this information from the raw files. \\
Existing network traffic classification methods include port-based identification, deep packet inspection (DPI), and statistical techniques using both machine learning and deep learning. While port-based identification\cite{zhao2021network} is computationally efficient, it only provides limited insight into application-level activity. DPI \cite{azab2022network} offers more detailed classification but comes with high computational costs, privacy concerns, and struggles with detecting zero-day threats. Supervised learning approaches can achieve high accuracy but depend heavily on well-labeled training datasets and careful feature engineering, whereas unsupervised methods \cite{blaise2020detection} reduce labeling efforts at the cost of precision. Semi-supervised techniques\cite{tongaonkar2008fast} combine the strengths of both, yet they increase computational complexity, and deep learning methods\cite{islam2023deep,sharmila2024maxvote}, despite eliminating manual feature engineering\cite{zhu2020you}, require vast datasets and substantial resources.
Thus, inspecting whether a packet is malicious, suspicious, or non-malicious with a threat intelligent platform like, Virus Total, helps organizations to detect and prevent threats, respond to incidents, protect users, maintain security hygiene, comply with regulations, enhance threat intelligence, and assess risks effectively.  The motivation is defended with table \ref{purposeList}.
\begin{table}[!htbp]
	\caption{Some important purposes for integrating VirusTotal with Network Forensic to investigate on RaaS}
	\label{purposeList}
		\begin{tabular}{|c|c|c|}
			\hline
			P.No & \textbf{Main Purpose} & \textbf{Achieving the purpose} \\
			\hline
			P1 & Threat Detection and Prevention & i)Identifying Malicious URLs ii)Blocking Threats 
			\\
			\hline
			P2 & Incident Response & i)Analyzing Incidents ii)Forensic Investigations 
			\\
			\hline
			P4
			&Phishing Protection  
			&i)Preventing Phishing Attacks 
			ii)User Education 
			\\ \hline
			P5
			&Maintaining Security Hygiene
			&i)Regular Monitoring
			ii)Updating Security Databases 
			\\ \hline
			P5
			&Compliance and Reporting
			&i)Regulatory Compliance
			ii)Reporting and Documentation 
			\\ \hline
			P6
			&Enhancing Threat Intelligence
			&i)Building Threat Intelligence
			ii)Sharing Intelligence
			\\ \hline 
			P7	& Risk Assessment
			& i)Evaluating Risks
			ii)Proactive Measures
			\\ \hline
		\end{tabular}
\end{table}
\begin{figure}[h]
	\centering
	\includegraphics[width=6in]{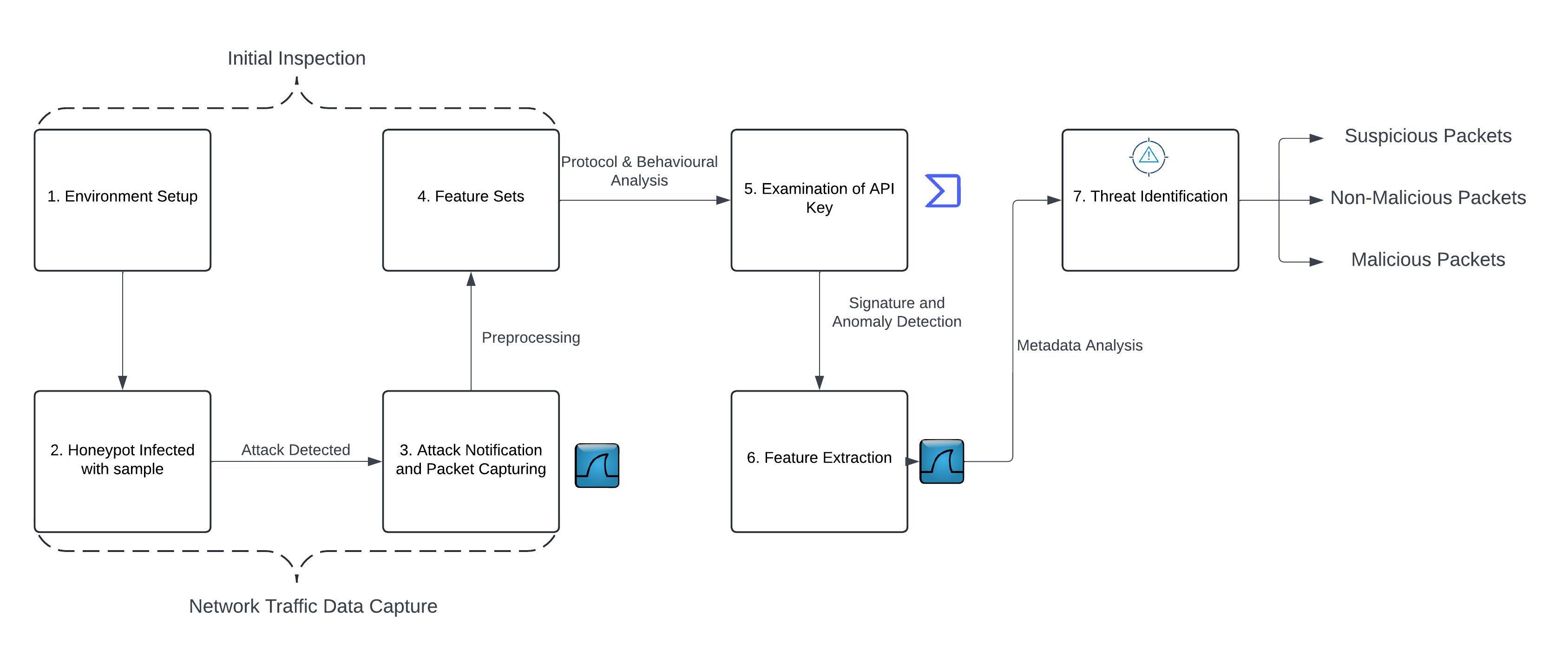} 
	\caption{Pipeline view of the proposed Solution with VirusTotal integration for packet classification.}
	\label{fig:RaasFlow}
\end{figure}
\section{Network Forensic Analysis of RaaS}
\label{Proposed}
Network data is volatile and differs from one incident to another incident. A strategic plan must be established to gather the attack vectors and Indicators of Compromise(IoC) for investigation. Network forensic analysis involves the monitoring and analysis of network traffic, events, logs, and communication patterns after cybersecurity incidents to gather information, secure legal evidence, and pinpoint the intrusion. 

Documenting a detailed report of the crime scene and duplicating all the collected digital shreds of the evidence will aid in efficient and accurate analysis of the evidence and to track all the visible data along with the metadata. Reporting all the shreds of evidence, reports, conclusions plays major significant role in the investigation.\\
VirusTotal is a threat intelligent platform associated with over 100 antivirus engines and a highly accurate classifier\cite{choo2023large} that correctly identify the attack types of malicious URLs at an early stage. These engines are from various vendors worldwide and they continuously contribute to the detection and analysis of malware and malicious IoCs. The list of these engines is subject to changes as new antivirus solutions are integrated into VirusTotal and existing ones change their names.  So the most up-to-date information can be checked from the official site of VirusTotal documentation.\\
In the proposed approach we investigate on network traffic captured for normal state and ransomware samples execution. Ryuk and Gandcrab are the chosen samples belonging to RaaS family. This approach aid in studying the behavior of RaaS family of malware. The results of the proposed method is verified by Virus Total API. The malicious traffic include more than 40\% malicious and suspicious packets over all packets of rate limit provided by Virus Total. Detailed description of the proposed solution is depicted in Figure. \ref{fig:RaasFlow}, further elaborated in this section. 

The method employed for feature extraction from the pcap file and VirusTotal API approach is described in Algorithm \ref{Algo2}. This Algorithm is designed to iterate over a list of URLs read from packets, make API requests to check their status, and categorize the URLs based on the API response using rate limit. The primary tasks it performs include rate-limiting the API requests, parsing JSON responses, and writing the results to different files based on whether the URLs are malicious or not.\\
%
\begin{algorithm}[h]
	\caption{Behavior analysis of packets with VirusTotal}
	\label{Algo2}
\small
	\begin{algorithmic}[1]
		\Require inputFile, ColIndex
		\Ensure outputFiles [cleanFile, malFile, suspFile]\\
		specify file paths for $cleanFile, malFile, suspFile$\\
		Set VirusTotal attributes [$apiKey$ and $URL$] with secure management\\
		validate Filepaths and column index
		extractColumn(inputFile, outputFile, colIndex)\\
		Get a list of URLs - Validate URL list\\
		Initialize counters:
		Start with $initialNum=1, reqMade=0, SNo=1$\\
		Record the $startTime$ \\
		Open $cleanFile, malFile, suspFile$ in append mode \\
		Handle file open errors
		Write $timestamp$ and $initialNum$ to files\\
		for each URL in the URL list\\
		\hspace{0.5cm} Validate URL and setting the VirusTotal attributes\\
		\hspace{0.5cm} if request rate limit is not reached\\
		\hspace{1cm}Set VirusTotal attributes\\
		\hspace{1cm}make API request \\
		\hspace{1cm}handling timeout and request failures\\ 
		\hspace{1cm}retry with exponential backoff\\
		\hspace{1cm}GetJSON response and handle JSON parsing errors\\
		\hspace{1cm}Get $timestamp$ for the current packetInfo\\
		\hspace{1cm}Write packet info with serial number to files \\
		\hspace{1cm}Update the $SNo$ by incrementing\\
		\hspace{0.5cm}else\\
		\hspace{1cm}wait and retry till log rate limit is hit\\
		Close all file and resources\\
		Return Log performance metrics and API usage statistics
	\end{algorithmic}
\end{algorithm}

\subsection{Environment setup with Honeypot }
A simulated environment is setup with virtualbox having Windows 10 and Kali Linux ensured with minimum of 8GB RAM and 200GB storage space, enables security for the host for secure examination of potentially dangerous network traffic and viruses. The tools we employed for network forensics are tcpdump, Wireshark, Network Miner, Brim, Snort, and Splunk. These are used to capture and analyze network traffic to index and analyze network data. Snort is used as an Intrusion Detection System (IDS) to invesigate on IoCs, also Tshark is used for metadata analysis.  

Malicious samples were collected from the malware repositories. After the Environment Setup, the installation of required tools and software, Ryuk  ransomware samples are executed in the vulnerable honeypot host machine mounted with network volume. 
Maintaining honeypot for a continuous period of 5 days, employing weak computers alongside the infected and inspected computers. The same experiment is repeated for Gandcrab sample. 

\subsection{PCAP capturing and preprocessing}
\small

Let $ \mathcal{D} $ represent the complete dataset used for the analysis. This dataset is composed of two disjoint subsets: a benign dataset $ \mathcal{D}_{benign} $, a collection of benign network traffic captures (PCAPs) obtained from public security repositories and research databases and a malicious dataset $ \mathcal{D}_{malicious} $, A collection of network traffic captures generated from controlled ransomware executions.
$ \mathcal{D}_{malicious} $ is generated by observing the execution of specific ransomware families.
Let $ M $ be the set of ransomware families studied, where $ M = \{ \text{Ryuk, Gandcrab} \} $.
Let $ T_{obs} $ be the observation period for each experiment, where $ T_{obs} = 5 $ days.
The experiment involves the infection of weak hosts, where the ransomware encrypts a shared directory $ S_d $.

For each malware $ m \in M $, network traffic was captured passively between the infected host and the server, resulting in a set of packets $ P_m $. The volume of captured packets is as follows:
Number of packets for Ryuk: $ |P_{\text{Ryuk}}| = 3,433,887 $.
Number of packets for Gandcrab: $ |P_{\text{Gandcrab}}| = 983,486 $.
Each packet capture $ P_m $ contains the network traffic corresponding to the file access operations performed on the shared directory $ S_d $.
The choice of passive network capture was made to minimize experimental interference.
Let $ C_{net} $ be the passive network traffic capture method and $ C_{host} $ be the local host-based I/O call capture method.
The interference $ \mathcal{I} $ with the host-server interaction using $ C_{net} $ is negligible: $ \mathcal{I}(C_{net}) \approx 0 $.
Conversely, the local host-based method results in a significant increase in CPU load: $ \Delta\text{CPU}_{load}(C_{host}) > 0 $.
Therefore, $ C_{net} $ was selected to preserve the natural behavior of the ransomware during execution. The resulting data is suitable for network forensic analysis.

\subsection{Featureset collection}
\small
The pcap file is analysed with Wireshark, to perform a quick scan to understand the general characteristics of the traffic, including packet sizes, protocols used, and any obvious anomalies. 
Suspicious patterns and traffic spikes indicating malicious activity is evident. Following this, Pre-processing ensued, involving the removal of additional non-required packets from the pcap dataset. Furthermore, feature sets were meticulously identified and integrated. These feature sets encompass various elements such as Protocol, HTTP Links, and Time-stamp, among others. Notably, the Time-stamp emerged as a pivotal parameter, pivotal in identifying attacks within the stipulated timestamp. 



Let $ P_m $ be the complete set of captured packets for a specific ransomware family $ m $, where $ m \in \{\text{Ryuk, Gandcrab}\} $. The initial analysis using Wireshark can be represented as a function that extracts high-level traffic characteristics.
Let $ \mathcal{A} $ be the analysis function.
$ \mathcal{A}(P_m) \rightarrow \{ S_m, \Pi_m, A_m \} $
where,
   $ S_m $ is the distribution of packet sizes in $ P_m $,
   $ \Pi_m $ is the set of network protocols identified in $ P_m $,
   $ A_m $ is a set of observed anomalies, such as suspicious patterns or traffic spikes. The presence of malicious activity, $ E_{malicious} $, is inferred if $ A_m \neq \emptyset $.
Following the initial analysis, a pre-processing stage is applied to refine the dataset. This involves filtering out irrelevant packets.
Let $ P_{raw} $ represent the raw packet capture for a given malware sample. Let $ P_{noise} $ be the set of identified non-required or noisy packets. The cleaned packet set, $ P'_m $, is obtained by the set difference:
$ P'_m = P_{raw} - P_{noise} $
This step ensures that subsequent analysis is performed only on packets relevant to the ransomware's activity.

From the cleaned packet set $ P'_m $, a feature set $ \mathcal{F} $ is defined for detailed analysis.
Let $ \mathcal{F} = \{ f_1, f_2, f_3, \dots, f_n \} $ be the set of $n$ features, this includes:
$ f_1 $: Protocol (e.g., TCP, UDP, HTTP).
$ f_2 $: HTTP Links.
$ f_3 $: Timestamp and so on.
For each packet $ p \in P'_m $, a feature extraction function, $ \text{Extract}(p, \mathcal{F}) $, generates a feature vector $ \mathbf{v}_p $:
$ \mathbf{v}_p = \langle f_1(p), f_2(p), f_3(p), \dots, f_n(p) \rangle $
The Timestamp, $ f_3(p) $, is noted as a critical feature for attack identification. The importance of a feature can be represented by a weight, $ w_i $. The high importance of the timestamp implies $ w_3 > w_i $ for $ i \neq 3 $.
The analysis also focuses on identifying specific malicious artifacts within the traffic of each ransomware family.
Let $ L_m $ be the set of HTTP links and $ C_m $ be the set of packets with unverified checksums for malware $ m $. The analysis yields:
For Gandcrab: $ L_{\text{Gandcrab}} $ and $ C_{\text{Gandcrab}} $.
For Ryuk: $ L_{\text{Ryuk}} $ and $ C_{\text{Ryuk}} $.
These sets of artifacts, $ L_m $ and $ C_m $, serve as direct evidence and are critical inputs for forensic analysis.

%
%
Malicious Packets often coincide with preceding packets temporally, accentuating the importance of time stamp during the detection process. Further protocol analysis, metadata analysis and behavioral analysis is performed to execute signature based detection and anomaly detection.

\subsection{Protocol and Behavior Analysis}

This step involves a deep inspection of individual network protocols to identify deviations from standard behavior.
Let $ P'_m $ be the pre-processed set of packets for a given malware $ m $. Let $ \Pi $ be the set of all protocols used in the traffic, where $ \Pi = $ \{ HTTP, DNS, SMTP, SMB, etc.\} .
For each protocol $ \pi \in \Pi $, we define a set of anomaly detection functions $ \Phi_{\pi} $. These functions operate on the subset of packets $ P'_{\pi, m} \subseteq P'_m $ that use protocol $ \pi $.
$ \Phi_{\pi}(P'_{\pi, m}) \rightarrow A_{\pi} $
where, $ A_{\pi} $ is a set of identified anomalies specific to protocol $ \pi $. The function $ \Phi_{\pi} $ checks for Abnormal packet structures, non-standard communication, encrypted traffic analysis. 
Abnormal packet structures are deviations in packet size, headers, or payload against protocol specifications.
Non-standard communication is unexpected communication patterns (e.g., unusual command sequences).
Encrypted traffic analysis is for encrypted traffic $ P'_{enc, m} \subseteq P'_m $, a decryption function $ \text{Decrypt}(p, K) $ is applied, where $ p \in P'_{enc, m} $ and $ K $ is a set of potential decryption keys or reverse engineering techniques. A successful decryption resulting in a malicious payload $ p_{malicious} $ is a strong indicator of an attack.

Ransomware threats leverage multiple protocols for payload delivery. Let $ V_{\pi} $ be a binary variable indicating the use of protocol $ \pi $ as an attack vector. The set of potential attack vectors is: $ \mathcal{V} = \{ \pi \in \Pi \mid V_{\pi}=1 \} $
Based on our analysis, this set commonly includes {SMTP, POP3, IMAP, HTTP, HTTPS, SMB, FTP}.
For behavioral analysis, our focus shifted from individual packets to higher-level network events and flows. The tool Zeek is used to transform raw packet data into structured logs of network behavior.

Let $ Z $ be the Zeek analysis function that processes the packet set $ P'_m $.
$ Z(P'_m) \rightarrow \mathcal{L}_{events} $ where, $ \mathcal{L}_{events} $ is a set of high-level event logs. From these logs, specific malicious behaviors $ B_i $ are identified. Let $ \mathcal{B} = \{ B_1, B_2, \dots, B_k \} $ be the set of monitored malicious behaviors, such as:
$ B_1 $: Port Scanning.   
$ B_2 $: Brute-force Login Attempts.
$ B_3 $: DNS Tunneling.
$ B_4 $: Command and Control (C2) Communication and so on.
The presence of a behavior $ B_i $ in the logs is a strong indicator of malicious activity.
Flow data is analyzed to understand the relationships and communication patterns between hosts. A network flow $ F_{ij} $ between host $ i $ and host $ j $ is characterized by a tuple:
$ F_{ij} = (\text{src\_ip}, \text{dst\_ip}, \text{protocol}, \text{port}, \text{frequency},  \text{duration}) $.
Analysis of these flows can reveal suspicious connections. Furthermore, specific protocols like IGMP are examined for misuse.

Let $ U_{IGMP} $ be the usage pattern of the IGMP protocol. If $ U_{IGMP} $ corresponds to known malicious patterns (e.g., multicast group-based propagation for espionage or ransomware distribution), it is flagged as an indicator of compromise.
Finally, IP addresses identified in suspicious flows are validated against threat intelligence platforms. Let $ \text{IP}_{suspicious} $ be a suspicious IP address. Its reputation score, $ \text{Rep}(\text{IP}_{suspicious}) $, is queried from the VirusTotal platform. An outcome where $ \text{Rep}(\text{IP}_{suspicious}) $ is inconclusive highlights the complexity and limitations of single-point threat detection methods.

\subsection{Signature Based Anamoly Detection}

This phase utilizes Intrusion Detection Systems (IDS) like Snort and Suricata to match network traffic against a database of known threats.
Let $ \mathcal{R} $ be the set of all rules (signatures) in the updated signature database, such that $ \mathcal{R} = \{r_1, r_2, \dots, r_n\} $. Let $ P'_m $ be the pre-processed packet capture for a given malware family $ m $.
The signature-based detection function, $ \Delta_{sig} $, is applied to each packet $ p \in P'_m $. An alert is generated if a packet's content matches any rule in the database.
$ \text{Alert}(p) = 1 \quad \text{if} \quad \exists r \in \mathcal{R} \text{ such that } p \text{ matches } r $.
$ \text{Alert}(p) = 0 \quad \text{otherwise} $.

Suricata enhances this by applying these rules to various protocols $ \pi \in $ \{HTTP, SMTP, FTP, DNS\}, allowing for the identification of protocol-specific threats. The live detection test can be modeled as a function that verifies the effectiveness of $ \mathcal{R} $ against a set of simulated intrusions $ I_{sim} $.
This process identifies deviations from established normal network behavior.
Let $ \mathcal{B} $ be a baseline model representing normal network behavior, defined by metrics such as traffic volume $ V $, protocol usage distribution $ \Pi_{dist} $, and source/destination pairs $ (\text{IP}_{src}, \text{IP}_{dst}) $.
An anomaly $ A $ is detected if the current traffic observation $ T_{obs} $ significantly deviates from the baseline $ \mathcal{B} $. This can be expressed as a function $ \delta $ that measures the distance or divergence between the observation and the baseline. An anomaly is flagged if this divergence exceeds a threshold $ \theta $.
$ A(T_{obs}, \mathcal{B}) = 1 \quad \text{if} \quad \delta(T_{obs}, \mathcal{B}) > \theta $

A Random Forest model, $ \mathcal{M}_{RF} $, is trained to classify behavior. Let $ \mathbf{X}_{train} $ be a feature matrix representing normal network behavior, where each row is a feature vector $ \mathbf{v}_p $ for a packet $ p $. The model learns a function $ f_{RF} $ that identifies outliers.
For a new, unseen packet with feature vector $ \mathbf{v}_{new} $, the model assigns an anomaly score $ s_{anom} $.
$ s_{anom} = f_{RF}(\mathbf{v}_{new}) $
The packet is classified as an anomaly if its score falls outside the range established for normal behavior.
This step integrates external threat intelligence and culminates in a packet integrity scoring system.
Let $ \mathcal{I} = \{\text{ip}_1, \text{ip}_2, \dots, \text{ip}_k\} $ be the set of Indicators of Compromise (IoCs) extracted from the network traffic. Each IoC is verified using the VirusTotal API, represented by the function $ \text{VT}(\text{ip}_i) $, which returns a reputation score or classification.

A scoring function, $ \text{Score}(p) $, is defined to evaluate the integrity of each packet $ p $. This score is a weighted sum of various features, where a higher score indicates a non-malicious packet.
Let $ \mathbf{v}_p = \langle f_1, f_2, \dots, f_j \rangle $ be the feature vector for packet $ p $. The score is calculated as:
$ \text{Score}(p) = \sum_{j=1}^{N} w_j \cdot g_j(f_j) $
where,
 $ w_j $ is the weight assigned to feature $ f_j $.
 $ g_j(f_j) $ is a function that maps the feature value to a component of the integrity score. For instance, if $ f_j $ is an IP address, $ g_j $ could be based on its VirusTotal reputation.
A packet is considered potentially malicious if its score falls below a defined integrity threshold $ \tau_{integrity} $.
$ \text{IsMalicious}(p) = 1 \quad \text{if} \quad \text{Score}(p) < \tau_{integrity} $

\subsection{Metadata Analysis}
This step focuses on extracting specific metadata fields from the packet capture $ P'_m $ for detailed analysis. Let $ \mathcal{E} $ be the metadata extraction process, which utilizes the command-line tool TShark.
The extraction of a specific metadata field $ f $ 
is performed by applying a TShark command, which can be represented as a function $ \text{TShark}(P'_m, f) $. This function filters the packet capture based on a specific criterion (e.g., protocol) and extracts the values of the desired field.
Let $ D_{dns} $, $ H_{http} $, and $ U_{agent} $ be the sets of extracted metadata:
1. DNS Requests: $ D_{dns} = \text{TShark}(P'_{\text{Ryuk}}, \text{``dns.qry.name"}) $
2. HTTP Hostnames: $ H_{http} = \text{TShark}(P'_{\text{Ryuk}}, \text{``http.host"}) $
3. HTTP User-Agents: $ U_{agent} = \text{TShark}(P'_{\text{Ryuk}}, \text{``http.user\_agent"}) $.
This process is executed within a Kali Linux environment, and the outputs are saved for subsequent analysis.

The extracted metadata is cross-referenced with threat intelligence feeds to identify known malicious indicators.
Let $ \mathcal{M} $ be the set of all extracted metadata, where $ \mathcal{M} = D_{dns} \cup H_{http} \cup U_{agent} $. Let $ \mathcal{I}_{known} $ be the set of known Indicators of Compromise (IoCs) from threat intelligence sources.
The correlation process identifies the intersection between the extracted metadata and the known IoCs:
$ \mathcal{I}_{found} = \mathcal{M} \cap \mathcal{I}_{known} $
where, $ \mathcal{I}_{found} $ is the set of confirmed malicious indicators found in the traffic. This includes known malware signatures, suspicious domains, and malicious IP addresses.


The experimental analysis yielded several key findings:\\
Malicious IP Identification: A subset of IP addresses, $ \text{IP}_{suspicious} \subset \mathcal{M} $, was identified as potentially harmful. 
Their reputation was confirmed using a threat intelligence platform like VirusTotal.
$ \text{Reputation}(\text{IP}_1) = \text{Malicious} $
$ \text{Reputation}(\text{IP}_2) = \text{Suspicious} $
Unverified Checksums: The analysis identified a set of packets, $ P_{bad\_checksum} \subset P'_m $, with unverified checksums, serving as a significant indicator of potential network anomalies or manipulation.\\
Protocol Usage: The protocols SSDP and DNS were identified as key vectors for ransomware dissemination and C2 communication. Let $ \Pi_{mal} = \{\text{SSDP, DNS}\} $ be the set of protocols highly associated with this malicious activity.\\
System Information Leakage: Analysis of TCP logs revealed that certain fields, such as the `NOTE' column, could leak system information, providing an unintended channel for reconnaissance.

Following the metadata analysis, objects from the pcap file are exported for further forensic investigation. This is performed in a separate virtualized environment to contain any potential threats.

\section{Results and Discussion}
\label{Results}
\begin{figure}[h]
	\centering
	\includegraphics[width=3in]{"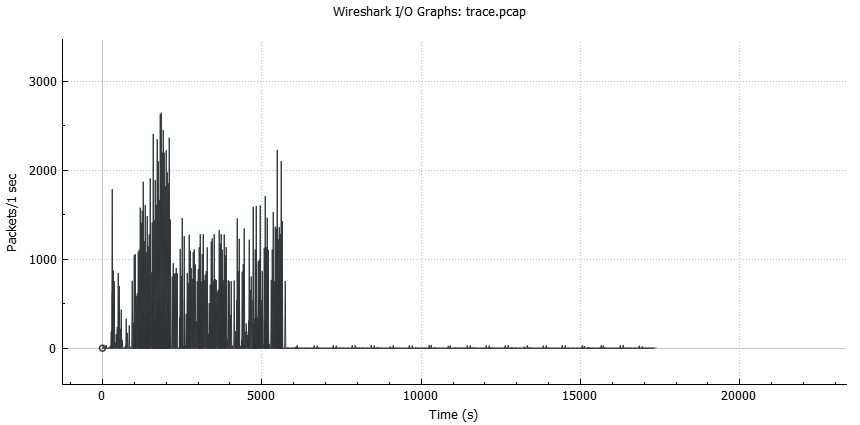"}
	\caption{Malicious activity observed in IO Graph for Gandcrab Pcap.}
	\label{fig:GandCrabIO}
	
\end{figure}

\begin{figure}[h]
	\centering
	\includegraphics[width=3in]{"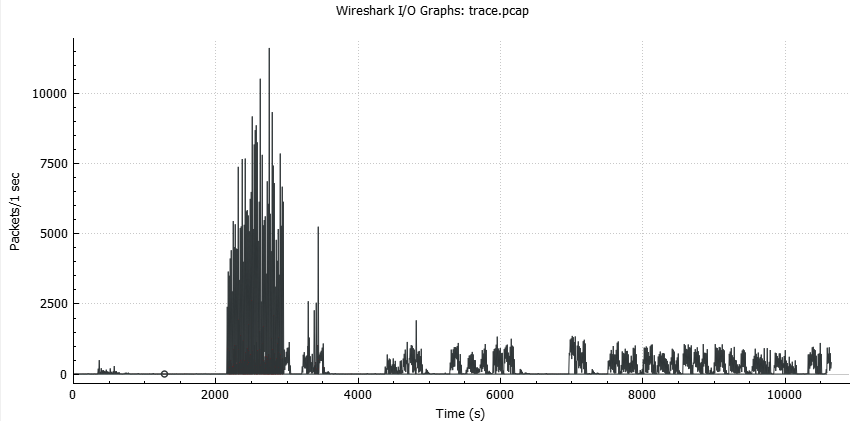"}
	\caption{Malicious activity observed in IO Graph for Ryuk Pcap}
	\label{fig:RyukIO}
	
\end{figure}

With the proposed approach of integrating VirusTotal with Network Forensics, more than 40\% of packets are found malicious in the RaaS pcaps. It serve as a major contribution for designing the packet signatures for RaaS group of malware. In this section we enumerate the packet statistics and IO graphs. Figure \ref{fig:GandCrabIO} and \ref{fig:RyukIO} depict the IO graphs derived from execution of Gandcrab and Ryuk respectively, showing malicious activity.  
The y-axis represents “Packets per second” (pkts/second). Peaks indicate high packet activity, while troughs represent low activity. The graph shows varying traffic intensity. Some periods have close to 0 pkts/second, while others reach just above 3000 pkts/second in Fig.\ref{fig:GandCrabIO} and above 7500 pkts/second in Fig.\ref{fig:RyukIO} indicating intense network activity. This is useful for identifying network performance issues, bottlenecks, and anomalies.

The peaks in the graph indicate periods of intense network activity. While high activity is expected, sustained peaks might lead to performance issues, if the network becomes saturated. The troughs represent moments of minimal packet activity. If these occur during critical operations, it could impact performance. The highest peaks around 3000 pkts/second in Fig.\ref{fig:GandCrabIO} and above 7500 pkts/second in Fig.\ref{fig:RyukIO} suggest potential congestion points with intense network activity. 
Bottlenecks may occur due to limited resources for example, bandwidth, processing power, memory etc., at specific points in the network. Sudden spikes or dips in packet activity are anomalies. These could be caused by unexpected events, such as a sudden burst of traffic or a device malfunction. The irregular pattern like sudden drops after sustained high activity warrants further investigation.

\begin{figure}
	\centering
	\includegraphics[width=3in]{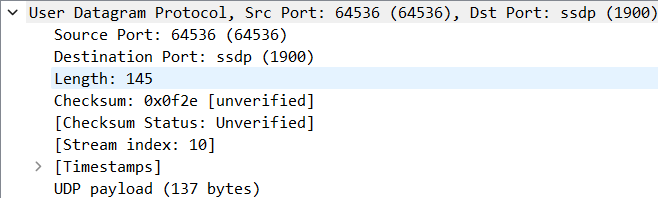}
	\caption{ Malicious Packet details identified with GandCrab: HTTP Link and unverified Checksum.}
	\label{fig:GandCrabimage1}
\end{figure}

\begin{figure}
	\centering
	\includegraphics[width=3in]{"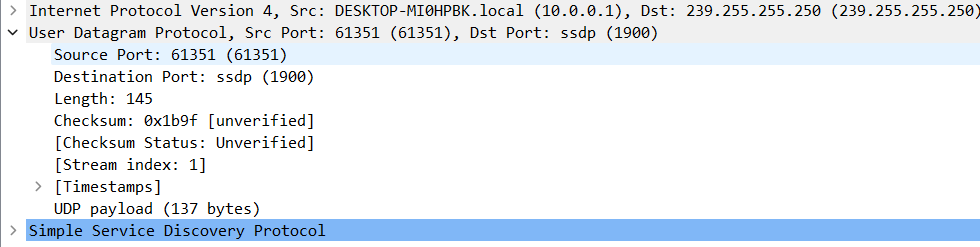"}
	\caption{Malicious Packet details identified with Ryuk: HTTP Link and unverified Checksum.}
	\label{fig:Ryukimage1}
	
\end{figure}

Packet captures are examined by the proposed model to identify the nature of the packets. HTTP links and unverified checksum details of the Gandcrab and Ryuk pcap are as shown in Figure \ref{fig:GandCrabimage1} and \ref {fig:Ryukimage1}. The details of malicious packets are tabulated in Table \ref{GandcrabPacketsTable} and \ref{RyukPacketTable} possessing five columns each representing a specific purpose in the analysis of Gandcrab and Ryuk. For brevity of space limited rows of output files($cleanFile, malFile, suspFile$) are shown here for exemplary purpose.\\
1.The Serial Number column denotes the sequential order of packets, facilitating easy reference and tracking.\\ 
2.The Date column records the precise date on which each packet was identified, providing temporal context to the analysis. \\
3.Timestamp column captures the exact time of packet identification, ensuring granularity in the temporal aspect of our investigation.\\
4.IoC extracted representing the details present in the packet under concern.\\
5.Classification marker, indicating whether the packet in question is classified as Malicious, Suspicious, or Non-Malicious\\
6.Additionally, the Information column complements the classification data by providing insights into the number of total sites checked to validate the accuracy of our classification results.\\

\begin{table}
	\caption{GandCrab Ransomware Packets Classification Results}
	\label{GandcrabPacketsTable}
	\resizebox{1.1\columnwidth}{!}{
		\begin{tabular}{|c|c|c|c|c|c|}
			\hline
			S.No & Date & Timestamp & IoC & Result & Information \\
			\hline
			\textit{53} & 2024-02-29 & 22:49:36 & 239.255.255.250 & Malicious & Detected by 4 Solutions \\
			\textit{104} & 2024-02-29 & 23:54:28 & 224.0.0.252 & Malicious & Detected by 2 Solutions \\
			\textit{5} & 2024-02-29 & 21:18:30 & ff02::1:ffb3:8513 & Suspected & Detected by 2 Solutions \\
			\textit{39} & 2024-02-29 & 22:18:45 & ff02::16 & Suspected & Detected by 4 Solutions \\
			\textit{2} & 2024-02-29 & 21:16:42 & 10.0.0.255 & NOT Malicious & Confirmed by 32 Solutions \\
			\textit{121} & 2024-03-01 & 00:16:03 & dns.google & NOT Malicious & Confirmed by 39 Solutions \\
			\hline
		\end{tabular}
	}
\end{table}

\begin{table}
	\caption{ Ryuk Ransomware Packets Classification Results}
	\label{RyukPacketTable}
	\resizebox{1.1\columnwidth}{!}{
		\begin{tabular}{|c|c|c|c|c|c|}
			\hline
			S.No & Date & Timestamp & IoC & Result & Information \\
			\hline
			\textit{4} & 2024-03-06 & 22:15:09 & igmp.mcast.net & Malicious & Detected by 1 Solutions \\
			\textit{25} & 2024-03-06 & 22:41:48 & 239.255.255.250 & Malicious & Detected by 5 Solutions \\
			\textit{14} & 2024-03-06 & 22:27:50 & DESKTOP-MI0HPBK.local & Suspected & Detected by 2 Solutions \\
			\textit{23} & 2024-03-06 & 22:39:16 & Broadcast & Suspected & Detected by 1 Solutions \\
			\textit{58} & 2024-03-06 & 23:23:38 & 224.0.0.252 & NOT Malicious & Confirmed by 38 Solutions \\
			\textit{62} & 2024-03-06 & 23:28:43 & mdns.mcast.net & NOT Malicious & Confirmed by 37 Solutions \\
			\hline
		\end{tabular}
	}
\end{table}
\begin{figure}[H] 
	\centering
	\includegraphics[width=2.5in]{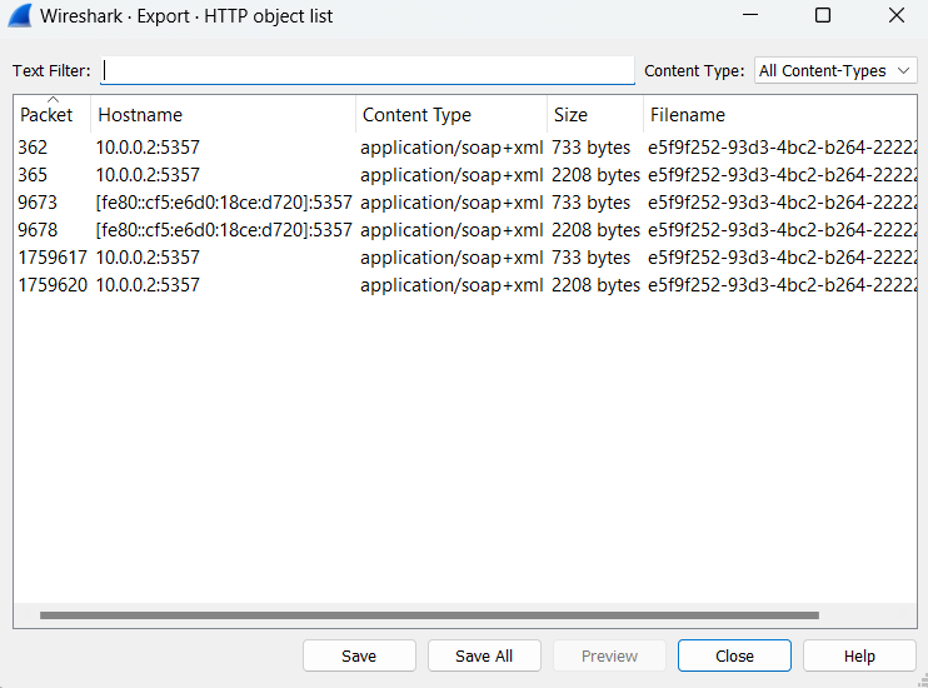}
	\caption{Six files associated with Ryuk analysed during network forensics}
	\label{fig:RyukFilesExporting}
\end{figure} 
Metadata analysis revealed about the hidden objects with the ramsomware. When Ryuk pcap is analysed in the virtualbox to export the associated objects, it is observed that it is associated with some hidden files. Ryuk is associated with 6 files as shown in Figure \ref{fig:RyukFilesExporting} and Gandcrab is also associated with 2 files. Further, these files were exported in isolated environment for reverse engineering analysis. It can be observed that the exported files are of type application. Reverse engineering these files will aid in discovering more about the connected attack vectors associated with the ransomware.
%
%
%
%
\section{Conclusion}
\label{Conclusion}
Ransomware as a Service is the greatest havoc these days with the game changing AI technology and blockchains. In this work we have conducted network forensic analysis of the packets captured during network traffic of two RaaS samples, Ryuk and Gandcrab and proposed a Virus Total integrated security approach for behavioral analysis of RaaS. 983486 packets were captured by execution of Gandcrab Ransomware. Similarly, 3433887 packets were captured by execution of Ryuk Ransomware. 
Investigating from the packets captured by the network traffic, we were successful in classifying the packets into malicious, non-malicious and suspicious. More than 40\% of packets in RaaS samples are identified as malicious. This is also verified and validated by VirusTotal API Approach.
The analysis can provide valuable threat intelligence on RaaS operators, including their infrastructure, tools, and methods.
The classified packets can be used to create custom signatures for intrusion detection systems and antivirus softwares, further can be extended for more number of RaaS samples.
The analysis can aid in digital forensics investigations by providing insights into the attack's timeline, data exfiltration, and encryption methods. The results can be used to improve security products, such as endpoint detection and response (EDR) solutions or next-generation firewalls (NGFWs). 
In future, we are working on automating the proposed approach with AI methods to ease the early detection of the upcoming attack by learning the inherent behavior of RaaS analysis. 

{\footnotesize 	
	\bibliographystyle{ieeetr}
	\small
	\bibliography{sample2} 
}

\noindent
%
%
%
%
%
%
%
%
\end{document}